\documentstyle[12pt]{article}

\begin{document}

\hfill{SPhT/94-098}
\vfill
\centerline{\bf THE FREE ENERGY OF HOT QED AT THREE AND A HALF
LOOPS \footnote{Presented at the Workshop of Quantum Infrared Physics,
Paris, June 6-10 1994.}}
\baselineskip=16pt
\vspace{0.7cm}
\centerline{\rm Rajesh R. Parwani\footnote{E-mail: parwani@wasa.saclay.cea.fr}}
\baselineskip=13pt
\centerline{\it Service de Physique Theorique, CE-Saclay }
\baselineskip=12pt
\centerline{\it F-91191, Gif-sur-Yvette, France }
\vspace{0.3cm}
\centerline{and}
\vspace{0.3cm}
\centerline{\rm Claudio Corian\`{o}\footnote{E-mail: coriano@hep.anl.gov}}
\baselineskip=13pt
\centerline{\it High Energy Physics Division, Argonne National Laboratory }
\baselineskip=12pt
\centerline{\it 9700 South Cass, Il 60439, USA.}

\vspace{0.95cm}
\centerline{\bf Abstract}
\vspace{0.15 cm}
{The computation of order $e^4$ and $e^5$ contributions to the
pressure of massless quantum electrodynamics at a temperature $T$ is
overviewed.}

\vspace*{-0.1 cm}
\baselineskip=14pt
\vspace{1.3 cm}

The thermodynamic properties of a QED plasma
may be determined from the partition function which can be
obtained perturbatively  using
techniques borrowed from field theory at zero temperature ($T$).
Indeed, the simplest
approach to calculating the partition function, or the free energy,
is to use the imaginary-time formalism whereby
the Feynman rules are as at $T=0$
but the energies take on discrete
Matsubara values.\\

However a naive application of these $T=0$ like Feynman rules
soon leads to the appearance of power-like infrared (IR)
singularities in diagrams. When these IR singularities
from an infinite set of diagrams are resummed, one obtains
an expansion in $\sqrt{e^2}$ rather than $e^2$, where $e$
is the QED coupling.
The best-known example of this phenomenon is the $e^3$ plasmon correction
to the free energy of QED first found by Gell-Mann and Brueckner
\cite{GB} in the
nonrelativistic context and later calculated relativistically \cite{AP}.
Physically, the appearance of these IR singularities and the
consequent breakdown of the naive perturbative expansion is
due to the Debye screening of electric fields
in a plasma. Unlike at $T=0$, the particles in the plasma are not free
(modulo ultraviolet renormalisations) but
perpertually under the influence of the surrounding particles.\\

In the language of field theory, the Debye screening manifests itself by the
nonvanishing limit $\Pi_{00}(p_0=0, \vec{p} \rightarrow 0)$ of the
electric polarization operator. To lowest order in the coupling, and in the
limit of high temperature, the electric screening mass-squared is given by a
one-loop calculation : $m^2 = \Pi^{1}(p_0=0, \vec{p} \rightarrow 0) =
e^2 T^2 /3$.  The presence of the large scale $T$ means that the
``loop correction'' need not be small compared to the bare propagator.
In imaginary time,
this is the case for the zero mode (Matsubara frequency) of the
electric propagator at small three momentum ($|\vec{p}| \sim eT$).
(By contrast, static magnetic fields are unscreened in a QED plasma.)\\

In general there are three ways that one can
account for large corrections, such as Debye screening,
in order to restore the perturbative expansion :

(i) Continue with bare Feynman rules and resum by hand
dangerous subsets of diagrams.
 This procedure is possible in simple cases but has the disadvantage that
one must carefully identify the relevant diagrams,
account for symmetry factors and prevent overcounting;
or

(ii) Begin with the non-perturbative skeleton-expansion and truncate down.
This is safe from the point of view of symmetry factors and overcounting
but is practical only in simple cases; or

(iii) Re-organise the bare Lagrangian by adding and subtracting the dominant
non-negligible effects (termed ``hard thermal loops'' by Braaten and
Pisarski \cite{BP}). This method has the virtue that one continues with
Feynman perturbation theory but with new effective propagators and vertices.\\

Usually, methods (i) and (ii) are  computationally efficient only for
static (zero external energy) Greens functions for which  all
the power counting analysis  can be done in
imaginary-time. For a non-static Greens function
one first requires its physical definition in real time (either by analytic
continuation from imaginary time or through
 the real-time formalism) and then the
method advocated in Ref.[3] is probably the most efficient. \\

Recently we computed the order $e^4$ ($3$-loop) contribution to the free
energy density
of massless QED at temperature $T$, going beyond the $e^3$ term known for
many years \cite{AP}. The method used was (i) with dimensional regularisation
being extensively employed to regulate various singularities
appearing at intermediate stages of the calculation.
{}From the technical point of view the fourth order calculation required the
evaluation of some complicated overlapping three-loop integrals that did not
appear in a similar three-loop calculation in $\phi^4$ theory \cite{FST}.\\

Following the order $e^4$ calculation in Ref.[5],
the $e^5$ ($3{1 \over 2}$ loop)
result was also obtained \cite{P}.
This latter term may be viewed as a correction to
the three-loop result as a consequence of Debye screening, just as the $e^3$
term is a similar correction to the two-loop (order $e^2$)
 result. Computationally,
the order $e^3$ calculation is {\em simpler} than the order $e^2$ calculation
because only the zero mode of the photon is involved and the loop integral
along that line becomes three-dimensional. Similarly, the
$e^5$ piece was easier to obtain than the $e^4$ piece, the sum of complicated
diagrams factorizing themselves into a product of simple one-loop integrals.
The $e^5$ contribution was also reconsidered from the point of view of
method (ii) in Ref.[7].\\

Remarkably, because the odd terms ($e^3$, $e^5$ ) are simple to calculate,
one is able to write a general identity \cite{P}.
Consider the gauge-invariant
contribution to the pressure
of { \em massless} QED, at temperature $T$, at order
$e^{2n}, (n \ge 1)$, coming from diagrams with one-fermion loop. Call this
contribution $P^{1F}_{2n}$. Then the order $e^{2n+3}$ piece is obtained by
dressing the photons of $P^{1F}_{2n}$ and is given by

\begin{equation}
P^{1F}_{2n+3} = {e^3 T^2 N^{1/2} \over 8 \pi \sqrt{3}} \
{\partial^{2} P_{2n}^{1F} \over \partial
\mu^2} |_{\mu=0} \; \; \; \, , n \ge 1 \, .
\label{spot}
\end{equation}
On the right-hand-side of Ref.(\ref{spot}), $\mu$ is the chemical
potential and $N$ is the number of massless electron flavours.
The generalisation of Eq.(\ref{spot}) to massive electrons at non-zero
chemical potential is given in Ref.[7].\\

Thus at least in this one case the ``IR problem'' of perturbation
theory at non-zero temperature has turned out to be a bonus in giving
us a cute relation like Eq.(\ref{spot}). Unfortunately the relation is not
quite useful since, as yet, we do not have a simple painless algorithm
to get the even pieces.\\

{\it Note added in proof :} At time of writing (August 1994),
the three-loop free-energy of hot Yang-Mills theory has been
calculated \cite{AZ}.\\

{\bf{Acknowledgements}}\\

R.P thanks Profs. H. Fried and A. White for the oppurtunity to present
these ``hot'' results at this workshop.\\

\newpage
\renewcommand{\theequation}{A.\arabic{equation}}
\setcounter{equation}{0}
\noindent
\noindent {\large\bf Appendix }
\vskip 3mm \noindent

The fine-structure constant at temperature
$T$ is $\alpha(T)=e^2(T)/4\pi$.
Defining $g^2 = \alpha(T) N / \pi$,  the pressure of QED with
$N$ massless Dirac fermions at nonzero temperature, $T$,
is then given by :

\begin{eqnarray}
P\over T^4 &=& a_0 + g^2 a_2 + g^3 a_3 + g^4 (a_4 + b_4 / N) +
g^5 (a_5 + b_5 / N) + O(g^6) \, ,
\end{eqnarray}

with

\begin{eqnarray}
a_0 &=& {\pi^2\over 45} \ (1 +{7\over 4}N) \, , \\
&& \nonumber \\
a_2 &=& - {5\pi^2\over 72} \, ,\\
&& \nonumber \\
a_3 &=&  {2 \pi^2 \over 9 \sqrt{3} } \, , \\
&& \nonumber \\
a_4 &=& - 0.757 \pm 0.004  \, , \\
&& \nonumber \\
b_4 &=&  0.658 \pm 0.006 \, , \\
&& \nonumber \\
a_5 &=&  { \pi^2 [1- \gamma - \ln(4 / \pi)] \over 9 \sqrt{3}} \
= \ 0.11473... \; , \\
&& \nonumber \\
b_5 &=&   {- \pi^2 \over 2 \sqrt{3} } \; .
\end{eqnarray}
\end{document}